\documentclass[%
 aip,
 amsmath,amssymb,
 reprint,%
]{revtex4-1}

\usepackage{graphicx}
\usepackage{dcolumn}
\usepackage{bm}
\usepackage[colorlinks, linkcolor=blue,anchorcolor=blue,citecolor=blue,urlcolor=blue]{hyperref}

\usepackage[utf8]{inputenc}
\usepackage[T1]{fontenc}
\usepackage{mathptmx}
\usepackage{etoolbox}
\usepackage{siunitx}

\newcommand{\re}{\mathrm{e}}
\newcommand{\ext}{\mathrm{ext}}
\newcommand\Rey{{\text{Re}}}  

\makeatletter
\def\@email#1#2{%
 \endgroup
 \patchcmd{\titleblock@produce}
  {\frontmatter@RRAPformat}
  {\frontmatter@RRAPformat{\produce@RRAP{*#1\href{mailto:#2}{#2}}}\frontmatter@RRAPformat}
  {}{}
}%
\makeatother
\begin{document}

\preprint{AIP/123-QED}

\title{Microscale Hydrodynamic Cloaking via Geometry Design in a Depth-Varying Hele-Shaw Cell}
\author{Hongyu Liu}
 \affiliation{ 
 Department of Mathematics, City University of Hong Kong, Kowloon, Hong Kong, China
}%
\author{Zhi-Qiang Miao}%
 \email{zhiqmiao@csu.edu.cn; zhiqmiao@cityu.edu.hk}
\affiliation{ 
 Department of Mathematics, City University of Hong Kong, Kowloon, Hong Kong, China
}%
\affiliation{School of Mathematics and Statistics, Central South University, Changsha, 410083, Hunan Province, China}

\author{Guang-Hui Zheng}
\affiliation{%
School of Mathematics, Hunan University, Changsha, 410082, Hunan Province, China
}%


\begin{abstract}
We theoretically and numerically demonstrate that hydrodynamic cloaking can be achieved by simply adjusting the geometric depth of a region surrounding an object in microscale flow, rendering the external flow field undisturbed. Using the depth-averaged model, we develop a theoretical framework based on analytical solutions for circular and confocal elliptical cloaks. For cloaks of arbitrary shape, we employ an optimization method to determine the optimal depth profile within the cloaking region. Furthermore, we propose a multi-object hydrodynamic cloak design incorporating neutral inclusion theory. All findings are validated numerically. The presented cloaks feature simpler structures than their metamaterial-based counterparts and offer straightforward fabrication, thus holding significant potential for microfluidic applications.
\end{abstract}
\maketitle

\section{Introduction} 
The pursuit of rendering objects invisible within fluid flows has long captivated researchers due to its significant potential for controlling and manipulating flows. Cloaking via metamaterials represents a classical approach toward this goal. Over the past 15 years, hydrodynamic cloaking has advanced rapidly, inspired by developments in metamaterial cloaks across diverse domains including ray optics~\cite{Leonhardt2006}, wave optics~\cite{Pendry2006,Alu2005}, microwave~\cite{Schurig2006}, infrared regimes~\cite{Valentine2009}, and related disciplines~\cite{Zhang2008,Farhat2009,Stenger2012,Yang2012,Narayana2012,Schittny2013,Deng2017,Deng2017full}.
Similar to the frequency-dependent nature of optical and acoustic cloaking~\cite{Cummer2008,Liu2009,Zhang2011}, hydrodynamic cloaking exhibits Reynolds number dependence. For different Reynolds number regimes, simplifications of the governing equations (Brinkman and Navier-Stokes equations) yield distinct versions requiring unique material or geometric parameters for the cloaking devices. At low Reynolds numbers, Urzhumov and Smith proposed theoretical frameworks for cloaking in porous media using anisotropic inhomogeneous permeability for three-dimensional unbounded flows~\cite{Urzhumov2011} and mixed positive-negative permeability for two-dimensional cylinder flows~\cite{Urzhumov2012}. Later, Park~\textit{et al.}~\cite{Park2019} experimentally realized hydrodynamic metamaterial cloaks in microchannels for creeping flows. At high Reynolds numbers, cloaking regions for water waves have been demonstrated~\cite{Farhat2008,Zou2019,Zhang2020}. However, these fluid-flow cloaks typically require spatially varying material parameters that demand complex metamaterials and pose fabrication challenges.

Recent interest has shifted toward metamaterial-free hydrodynamic cloaking. Boyko~\textit{et al.}~\cite{Boyko2021} introduced a novel electro-osmosis approach for microscale hydrodynamic cloaking and shielding in Hele-Shaw cells without metamaterials. Building on this foundation, we established rigorous mathematical frameworks~\cite{Liu2024,Liu2024enhanced,Kong2024} and subsequently demonstrated simultaneous microscale electric and hydrodynamic cloaking in Hele-Shaw configurations~\cite{Liu2024sim}, addressing the prior neglect of the electric field cloaking. These studies assumed uniform microchannel depth, thereby overlooking the influence of depth variations. 
In fact, Tay~\textit{et al.}~\cite{Tay2021} implemented a metamaterial-free hydrodynamic cloak via depth variation by transforming density-based cloaking formulas to depth-based equivalents using mass conservation. However, our analytical solutions for the depth-averaged model reveal a slight discrepancy in the cloaking formula, potentially attributable to their neglect of the velocity potential's quadratic dependence on channel depth. Replacing mass conservation with momentum conservation aligns their results with ours—a finding consistent with recent work on invisible hydrodynamic tweezers~\cite{Zhou2024}. We emphasize that Tay~\textit{et al.}'s foundational work remains highly instructive and inspired our current study.

In the present study, we exploit the strong dependence of microfluidic flow fields on geometric structure by utilizing non-uniform channel depth to simplify hydrodynamic cloak design, eliminating the need for electro-osmotic flow. Depth-varying geometries offer fabrication advantages through 3D lithography~\cite{He2016} in microfluidic applications. Notably, depth engineering constitutes an established technique for flow manipulation, evidenced by metamaterial-based water wave control~\cite{Chen2009,Zhao2021,Hua2022,Han2022}.
Here, we present a new theoretical method and numerical demonstration of metamaterial-free hydrodynamic cloaking in depth-varying microfluidic chambers. Under the lubrication approximation and in shallow geometries, the governing equation for Hele-Shaw flow reduces to Laplace's equation. Using a scattering cancellation approach, we develop a theoretical framework to derive analytical and optimal cloaking conditions for diverse geometrical shapes, extend these cloaking conditions to multi-object cloaking, and numerically validate these findings. 

\begin{figure*} 
\centering
\includegraphics[width=0.95\textwidth]{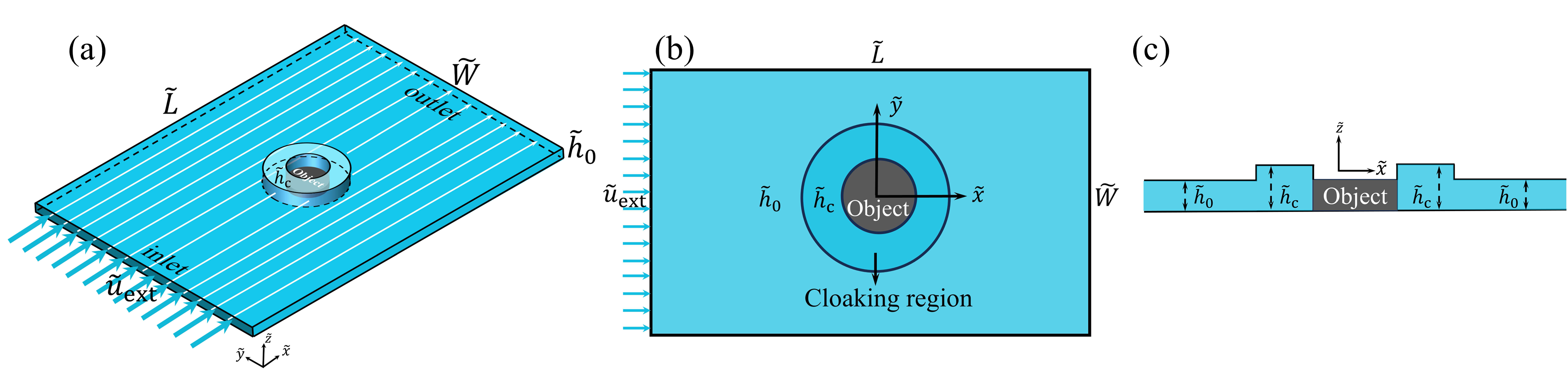}
\caption{\label{fig:schematic}Schematic of a depth-varying Hele-Shaw cell. (a) Microfluidic chamber with a circular cylinder confined between parallel plates. A depth-varying region on the top plate surrounds the object to enable cloaking. (b) Top view. (c) Side view. Flow outside the cloaking region remains uniform when cloaking is achieved.}
\end{figure*}


\section{Theoretical design and methods}

Figure \ref{fig:schematic} illustrates the configuration used for modeling and numerical simulation. 
The setup consists of a circular cylinder object with characteristic dimension $\tilde{r}_0$ confined between two parallel plates of length $\tilde{L}$ and width $\tilde{W}$, separated by a small variable gap $\tilde{h}$ ($\tilde{h} \ll \tilde{L}, \tilde{W}$). Assuming $\tilde{h} = \tilde{h}_c$ within the cloaking region and $\tilde{h}_0$ elsewhere, the configuration forms a depth-varying Hele-Shaw cell [Fig.~\ref{fig:schematic}(a)]. 
We emphasize that the cloaked object is not limited to circular cylinders; the approach applies to pillar-like objects with arbitrary cross-sectional geometries.
The liquid within the chamber undergoes pressure-driven flow with far-field mean velocity $\tilde{u}_{\text{ext}}$ along the $\tilde{x}$-axis. The object is surrounded by a cloaking region where the depth $\tilde{h}_c$ can be tuned relative to the background depth $\tilde{h}_0$. 
By controlling the depth within this cloaking region, the pressure distribution can be adjusted to conceal the object, rendering the external flow field indistinguishable from the background flow.

Considering shallow geometries ($\tilde{h} \ll \tilde{r}_0 \ll \tilde{L}, \tilde{W}$) and negligible fluid inertia, the governing equations for the depth-averaged velocity $\langle \tilde{\bm{u}}_{\parallel}\rangle$ and pressure $\tilde{p}$ in the $\tilde{x}$-$\tilde{y}$ plane are (see Appendix~\ref{sec:app1}):
\begin{equation}\label{eq:continuity}
    \tilde{\nabla}_{\parallel} \cdot (\tilde{h} \langle \tilde{\bm{u}}_{\parallel}\rangle ) = 0,
\end{equation}
\begin{equation}\label{eq:velocity}
    \langle \tilde{\bm{u}}_{\parallel}\rangle = 
    -\frac{\tilde{h}^2}{12\tilde{\mu}}\tilde{\nabla}_{\parallel}\tilde{p},   
\end{equation}
yielding
\begin{equation}\label{eq:pressure}
   \tilde{\nabla}_{\parallel} \cdot \left(\frac{\tilde{h}^3}{12\tilde{\mu}}\tilde{\nabla}_{\parallel}\tilde{p}\right) = 0.
\end{equation}
Here, $\tilde{\nabla}_{\parallel}$ and $\tilde{\nabla}_{\parallel} \cdot$ denote the two-dimensional gradient and divergence operators in the $\tilde{x}$-$\tilde{y}$ plane, respectively.

We impose the following boundary conditions to solve Eqs.~\eqref{eq:continuity}--\eqref{eq:pressure}. The no-penetration condition on the object surface requires
\begin{equation}\label{eq:penetration-condition}
    (\tilde{h}_0 \langle \tilde{\bm{u}}_{\parallel}\rangle) \cdot \hat{\bm{n}} = 0.
\end{equation}
At the interface between the cloaking region and the background flow, the matching conditions are
\begin{equation}\label{eq:matching-conditions}
   \tilde{p}_{\mathrm{in}} = \tilde{p}_{\mathrm{out}} \quad \text{and} \quad (\tilde{h}_0 \langle \tilde{\bm{u}}_{\parallel}\rangle_{\mathrm{in}}) \cdot \hat{\bm{n}} = (\tilde{h}_c \langle \tilde{\bm{u}}_{\parallel}\rangle_{\mathrm{out}}) \cdot \hat{\bm{n}},
\end{equation}
where the subscripts "in" and "out" denote fields inside and outside the cloaking region, respectively.

Given that the chamber boundaries are distant from the cloaking region, we model the domain as unbounded. The flow is driven by a uniform background flow $\tilde{u}_{\text{ext}}$ in the $\tilde{x}$-direction, yielding the far-field condition
\begin{equation}\label{eq:far-field}
   \langle \tilde{\bm{u}}_{\parallel} \rangle = \tilde{u}_{\text{ext}}\hat{\bm{x}} \quad \text{and} \quad \tilde{p} = -\frac{12\tilde{\mu}}{\tilde{h}_0^2}\tilde{u}_{\text{ext}}\tilde{x} \quad \text{as} \quad \tilde{r} \rightarrow \infty,
\end{equation}
where $\hat{\bm{x}} = (1,0)$ and $\tilde{r} = \sqrt{\tilde{x}^2 + \tilde{y}^2}$.

To cloak the object hydrodynamically, we require the flow field outside the cloaking region to attain a uniform value $\tilde{u}_{\text{ext}}$ along $\tilde{x}$-axis,
\begin{align}\label{eq:cloaking-condition-u}
\langle \tilde{\bm{u}}_{\parallel} \rangle = \tilde{u}_{\text{ext}}\hat{\bm{x}}.
\end{align}
Since the pressure and velocity in the outer region satisfy $\langle \tilde{\bm{u}}_{\parallel} \rangle = -\tilde{h}_0^2\tilde{\nabla}_{\parallel}\tilde{p}/(12\tilde{\mu})$, the condition \eqref{eq:cloaking-condition-u} can be expressed in terms of the pressure as
\begin{equation}\label{eq:cloaking-condition}
  \tilde{p} = -\frac{12\tilde{\mu}}{\tilde{h}_0^2}\tilde{u}_{\text{ext}}\tilde{x}
\end{equation} 
outside the cloaking region.

\subsection{\label{Analytical-method}Analytical method for circular and confocal elliptical cloaks}
We first design a circular cloak to achieve hydrodynamic cloaking for a cylindrical object using analytical solutions. Let the object radius be $\tilde{r}_i = \tilde{r}_0$ and the cloaking region radius be $\tilde{r}_e$. By solving the governing equation \eqref{eq:pressure} with boundary conditions \eqref{eq:penetration-condition}--\eqref{eq:far-field} in polar coordinates (see Appendix~\ref{sec:app2}), we obtain the pressure distribution. The cloaking condition \eqref{eq:cloaking-condition} is then satisfied when
\begin{equation}\label{eq:cloaking-condition-circle}
    \tilde{h}_c = \sqrt[3]{\frac{\tilde{r}_e^2 + \tilde{r}_i^2}{\tilde{r}_e^2 - \tilde{r}_i^2}}\,\tilde{h}_0.
\end{equation}
Note that Eq.~\eqref{eq:cloaking-condition-circle} differs from the corresponding cloaking formula in Tay \emph{et al.} \cite{Tay2021}, as their result omitted the cube root.

To generalize the analysis, we consider the same configuration shown in Fig.~\ref{fig:schematic}, replacing circular geometries with confocal ellipses. We next design an elliptical hydrodynamic cloak where an elliptical object is enclosed by a confocal elliptical cloaking region. Introducing elliptical coordinates $(\xi,\eta)$, the Cartesian coordinates $(\tilde{x},\tilde{y})$ transform as
 \begin{equation*}
     \tilde{x} = \tilde{r}_0 \cosh \xi \cos \eta \quad \text{and} \quad \tilde{y} = \tilde{r}_0 \sinh \xi \sin \eta,
 \end{equation*}
where $2\tilde{r}_0$ represents the focal distance. 

Denoting the inner and outer boundaries of the cloaking region by $\xi_i$ and $\xi_e$ respectively, and noting these confocal ellipses share common foci, the cloaking condition \eqref{eq:cloaking-condition} is satisfied when
\begin{equation}\label{eq:cloaking-condition-ellipse}
    \tilde{h}_c = \sqrt[3]{\tanh{\xi_e} \coth(\xi_e - \xi_i)}\,\tilde{h}_0.
\end{equation}
The complete theoretical derivation is provided in Appendix~\ref{sec:app2}.

\subsection{\label{Optimization-method}Optimization method for arbitrary-shaped cloaks}
Analytical solutions to the Laplace equation are generally limited to specific geometric shapes in specialized coordinate systems. Due to this limitation, hydrodynamic cloaks with arbitrary cross-sectional shapes cannot be designed using analytical methods. To enable cloaking for arbitrarily shaped objects, we propose an optimization framework to determine the optimal depth profile of the cloaking region, thereby achieving hydrodynamic cloaking performance that closely approximates that of an ideal cloak.

Let $\Omega$ denote the hydrodynamic cloak domain, with interior and exterior boundaries $\partial \Omega_i$ and $\partial \Omega_e$, respectively. According to the matching conditions \eqref{eq:matching-conditions} and the cloaking criterion \eqref{eq:cloaking-condition}, perfect cloaking requires that the following overdetermined boundary value problem admits a unique solution
\begin{align*}
    \tilde{\Delta} \tilde{p} &= 0 
        && \text{in } \Omega, \\
    \tilde{\nabla}_{\parallel} \tilde{p} \cdot \hat{\bm{n}} &= 0 
        && \text{on } \partial \Omega_i, \\
    \tilde{p} &= \tilde{p}_b 
        && \text{on } \partial \Omega_e, \\
    \tilde{h}_0^3 \tilde{\nabla}_{\parallel} \tilde{p}_b \cdot \hat{\bm{n}} &= \tilde{h}_c^3 \tilde{\nabla}_{\parallel} \tilde{p} \cdot \hat{\bm{n}} 
        && \text{on } \partial \Omega_e,
\end{align*}
where $\tilde{p}_b = -12\tilde{\mu}\tilde{u}_{\mathrm{ext}}\tilde{x}/\tilde{h}_0^2$ denotes the background pressure field.

For arbitrary geometries, the existence of such solutions cannot be guaranteed analytically. Consequently, we reformulate the cloak design problem as a constrained nonlinear optimization
\begin{equation*}
    \min_{\tilde{h}_c} \int_{\partial \Omega_e} 
    \left| \tilde{h}_0^3 \tilde{\nabla}_{\parallel} \tilde{p}_b \cdot \hat{\bm{n}} 
    - \tilde{h}_c^3 \tilde{\nabla}_{\parallel} \tilde{p} \cdot \hat{\bm{n}} \right|^2 
    \mathrm{d}s,
\end{equation*}
subject to $\tilde{h}_c > \tilde{h}_0$, where $\tilde{p}$ satisfies
\begin{align*}
    \tilde{\Delta} \tilde{p} &= 0 
        && \text{in } \Omega, \\
    \tilde{\nabla}_{\parallel} \tilde{p} \cdot \hat{\bm{n}} &= 0 
        && \text{on } \partial \Omega_i, \\
    \tilde{p} &= \tilde{p}_b 
        && \text{on } \partial \Omega_e.
\end{align*}

Analogous to \cite{Liu2024}, the existence, uniqueness, and stability of minimizers for this constrained optimization problem can be rigorously established. For numerical implementation, Newton-type methods are suitable for locating the optimal depth $\tilde{h}_c$, as the cost functional depends exclusively on this parameter.

\subsection{Multi-object cloak}
Thus far, our theoretical analysis has focused on hydrodynamic cloak design for single objects. However, cloaking multiple objects is of greater significance in practical physics and engineering applications. To address this challenge, we propose a multi-object hydrodynamic cloak by integrating neutral inclusion theory \cite{Zhou2006,He2013,Wang2025} with the analytical and optimization methods developed in Subsections \ref{Analytical-method} and \ref{Optimization-method}.
Neutral inclusion theory establishes that when objects are immersed in a free stream, mismatched flow boundary conditions near the objects generate flow disturbances. These disturbances are predominantly governed by the objects' volume (or surface area in two-dimensional configurations). Consequently, we adopt the principle that flow disturbances are primarily determined by the total invasive volume as our design foundation. This principle enables effective splitting of a single object into multiple constituents while preserving the overall flow disturbance characteristics.
Through this volume-conserving splitting approach, we can utilize the single-object hydrodynamic cloaks derived via the analytical and optimization methods in Subsections \ref{Analytical-method} and \ref{Optimization-method} to achieve multi-object hydrodynamic cloaking.

\section{Numerical Results and Discussion}
To demonstrate hydrodynamic cloaking, we perform three-dimensional finite-element simulations using COMSOL Multiphysics. The microfluidic chamber, filled with fluid of viscosity \SI{e-3}{Pa.s}, has dimensions of \SI{2}{\milli m} (length), \SI{1.4}{\milli m} (width), and \SI{15}{\micro m} (depth). An object with characteristic dimension $\tilde{r}_0 = \SI{100}{\micro m}$ is positioned at the chamber center, surrounded by a cloaking region with optimized depth. Pressure-driven flow is controlled by imposing zero pressure at the outlet and a normal inflow velocity $\tilde{u}_{\mathrm{ext}} = \SI{51}{\micro m/s}$ at the inlet. Physical and geometrical parameters match those in \cite{Boyko2021} to facilitate direct comparison.

\begin{figure}
\includegraphics[scale=1.08]{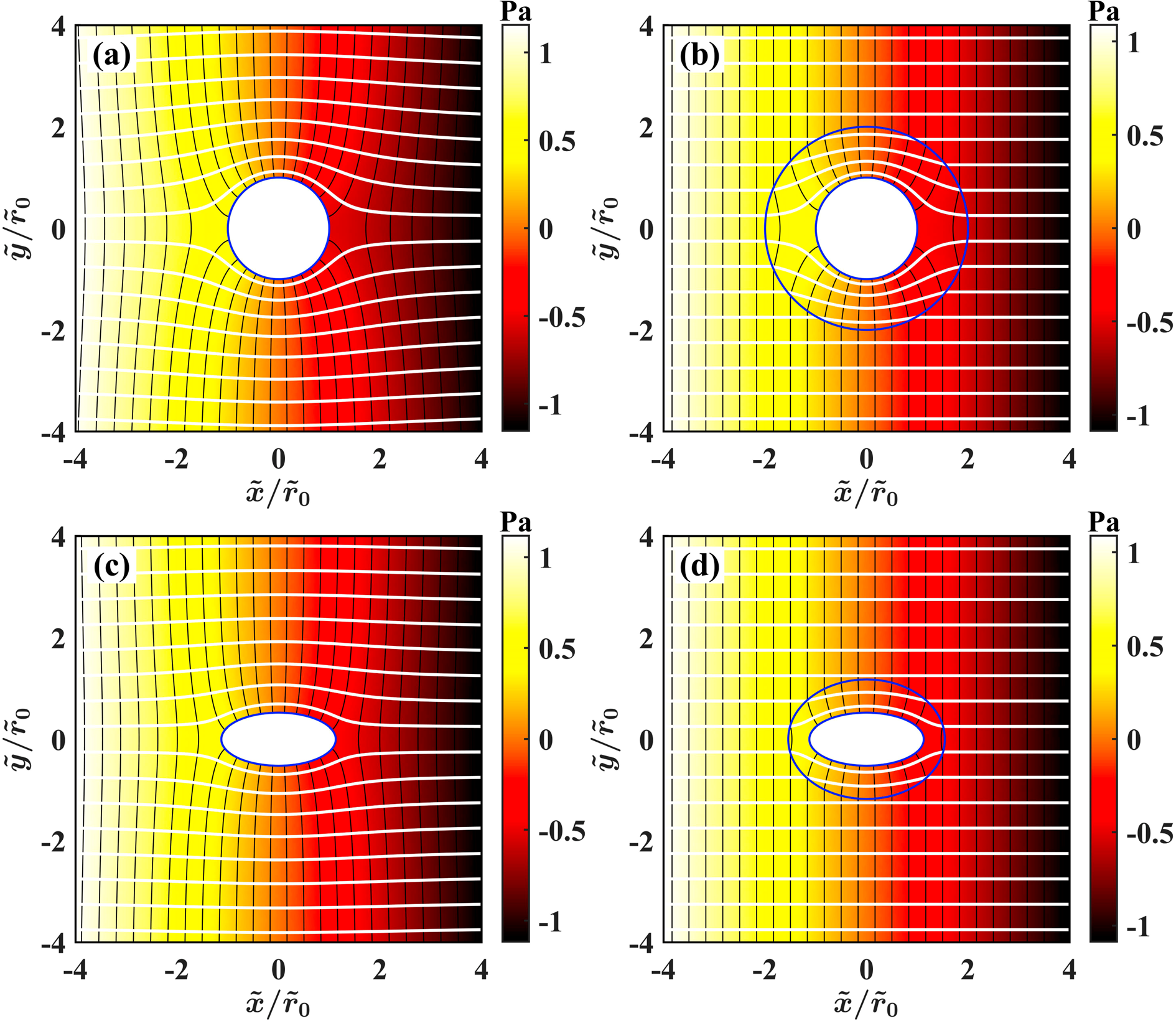}
\caption{\label{fig:circle-ellipse}Perfect hydrodynamic cloaking for cylindrical and confocal elliptical objects. (a--d) Pressure distribution (colormap), isobars (black lines), and streamlines (white lines) for (a,c) uncloaked and (b,d) cloaked configurations. The cloaking region (blue outline) is a circular/elliptical annulus with $\tilde{r}_e = 2\tilde{r}_i = 2\tilde{r}_0$ and $\xi_e = 2\xi_i = 1$, with depths of \SI{17.78}{\micro m} and \SI{17.72}{\micro m} respectively.}
\end{figure}

\begin{figure}
\includegraphics[scale=1.08]{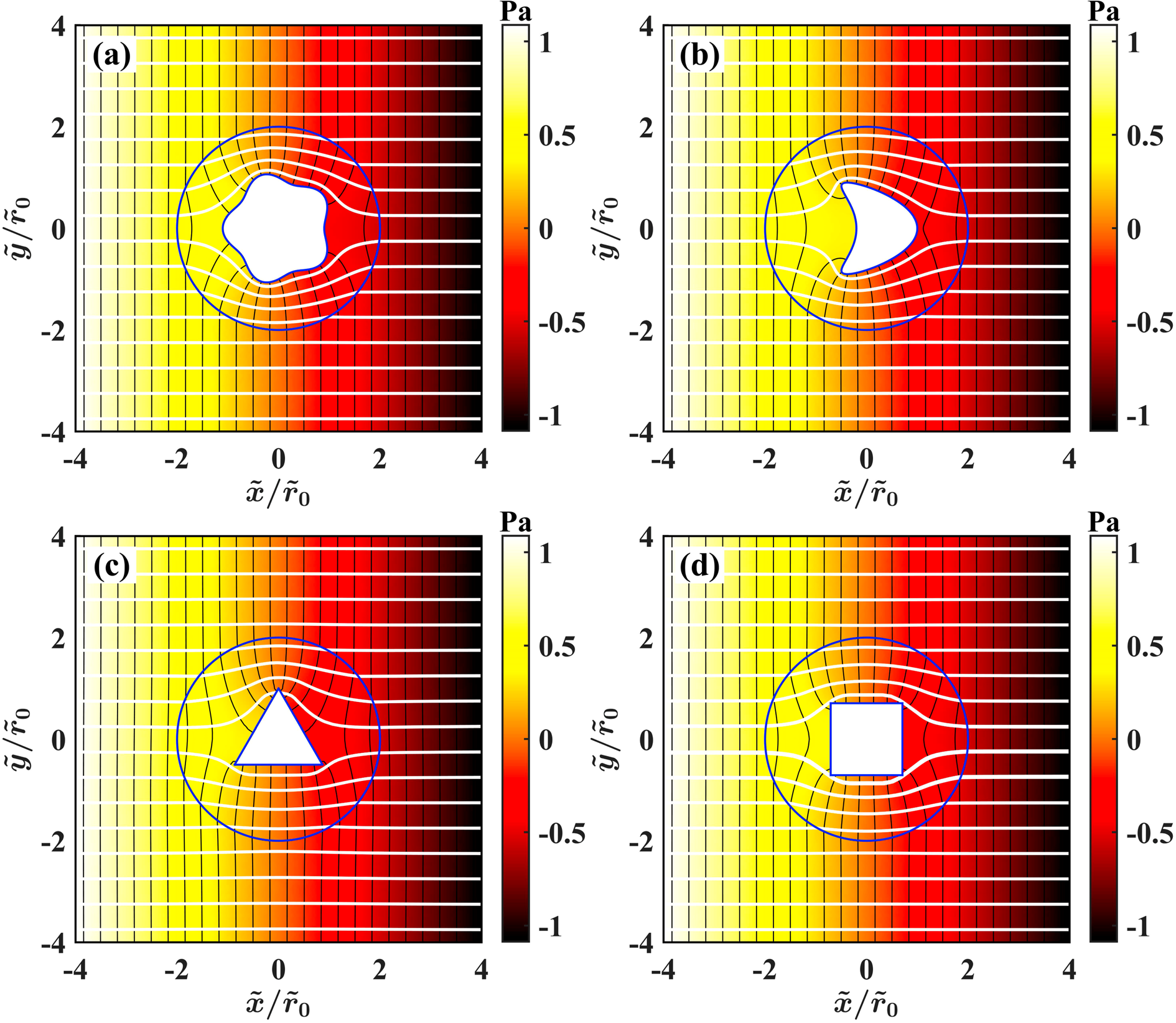}
\caption{\label{fig:arbitrary-shapes}Approximate cloaking for arbitrary-shaped objects: (a,b) regular (flower, kite) and (c,d) irregular (triangle, square) geometries. Cloaking regions (blue outlines) are circular with $\tilde{r}_e = 2\tilde{r}_0$, having optimal depths of \SI{17.92}{\micro m}, \SI{16.97}{\micro m}, \SI{16.45}{\micro m}, and \SI{16.85}{\micro m} respectively.}
\end{figure}
\begin{figure}
\includegraphics[scale=1.08]{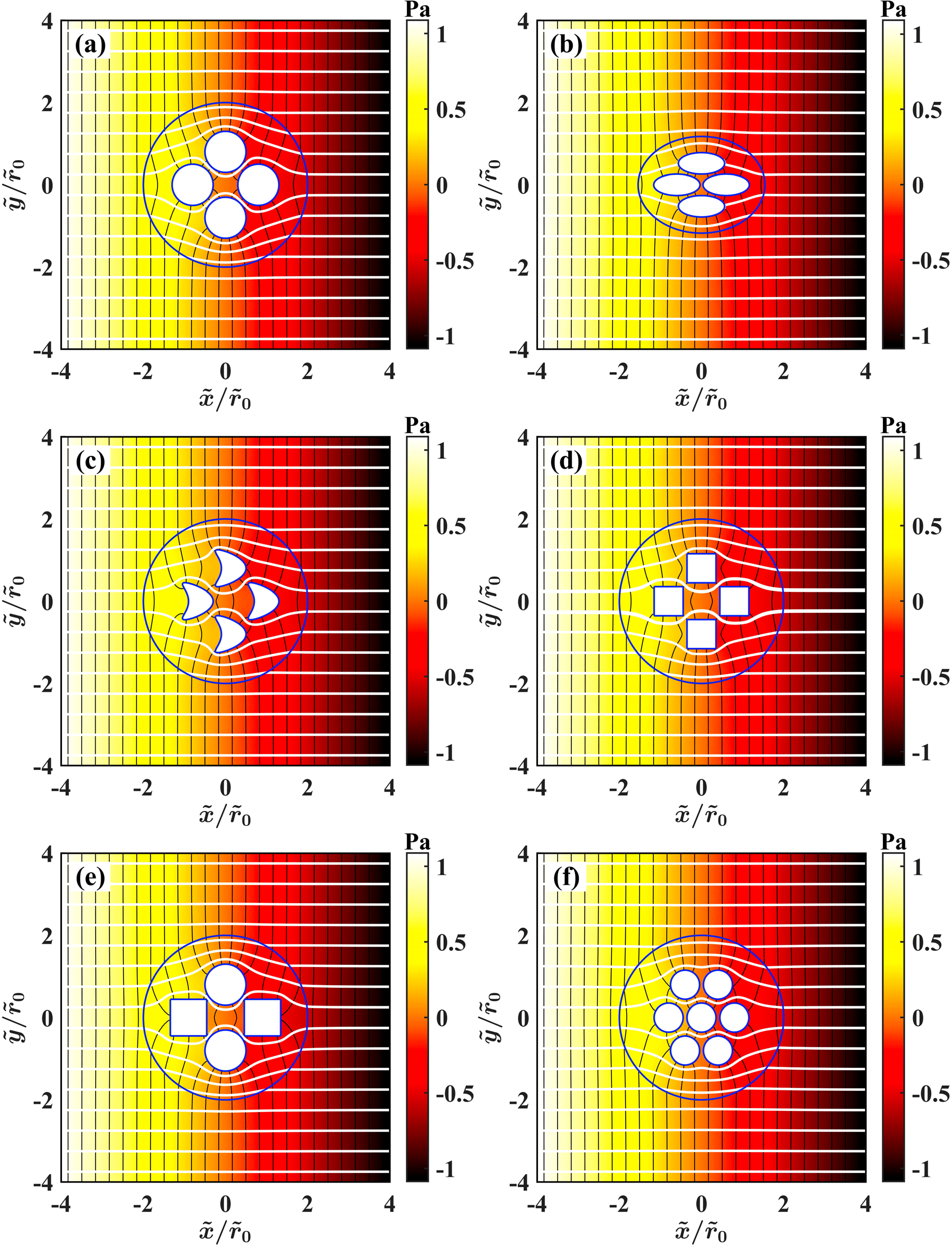}
\caption{\label{fig:multiple-objects}Approximate multi-object cloaking for: (a,b) four circular/elliptical cylinders, (c) four kites, (d) four squares, (e) mixed circles/squares, and (f) eight circles. Parameters match Figures~\ref{fig:circle-ellipse} and~\ref{fig:arbitrary-shapes}.}
\end{figure}

Figure~\ref{fig:circle-ellipse} shows pressure distributions (colormap), isobars (black lines), and streamlines (white lines) for circular and confocal elliptical cylinders in a Hele-Shaw cell under pressure-driven flow. Comparison between uncloaked [Figs.~\ref{fig:circle-ellipse}(a), \ref{fig:circle-ellipse}(c)] and cloaked [Figs.~\ref{fig:circle-ellipse}(b), \ref{fig:circle-ellipse}(d)] configurations reveals that under the respective cloaking conditions \eqref{eq:cloaking-condition-circle} and \eqref{eq:cloaking-condition-ellipse}, exterior isobars and streamlines remain straight and undisturbed relative to the uniform far field. These results validate the perfect cloaking performance achieved through analytical methods.

Figure~\ref{fig:arbitrary-shapes} demonstrates hydrodynamic cloaking for arbitrary-shaped objects surrounded by circular cloaking regions. Optimal cloaking conditions yield approximate cloaking for both regular [Figs.~\ref{fig:arbitrary-shapes}(a), \ref{fig:arbitrary-shapes}(b)] and irregular [Figs.~\ref{fig:arbitrary-shapes}(c), \ref{fig:arbitrary-shapes}(d)] geometries, confirming the optimization method's effectiveness for complex shapes.
Figure~\ref{fig:multiple-objects} extends these results to multi-object configurations. Perfect cloaking is observed for four circular/elliptical cylinders [Figs.~\ref{fig:multiple-objects}(a), \ref{fig:multiple-objects}(b)], while approximate cloaking persists for complex shapes [Figs.~\ref{fig:multiple-objects}(c), \ref{fig:multiple-objects}(d)], mixed geometries [Fig.~\ref{fig:multiple-objects}(e)], and larger object counts [Fig.~\ref{fig:multiple-objects}(f)].

In summary, our three-dimensional finite-element simulations validate the performance of analytically designed circular/confocal elliptical cloaks and optimization-based complex-shaped cloaks, showing excellent agreement with theoretical predictions.

\section{Conclusions}
We present a novel hydrodynamic cloaking method for microscale flows that exploits microfluidic chamber geometry, thereby eliminating the need for complex metamaterials. A complete theoretical framework has been developed to compute cloaking conditions for both simple and complex geometries. Furthermore, we demonstrate multi-object cloaking capability through strategic decomposition of a single object into smaller constituents. Numerical validation confirms that the proposed cloaks achieve effective hydrodynamic concealment across diverse object shapes and quantities.
Compared to existing electro-osmosis-based microscale cloaking approaches, this method establishes a simpler paradigm. While hydrodynamic shielding concepts are not addressed herein, as they require control-region depths that would violate depth-averaged model assumptions. Nevertheless, we note that increasing channel depth in the control region can reduce hydrodynamic forces on objects.
These findings open promising avenues for fluidic invisibility technologies with potential applications in microfluidics and biofluidics. Future work will explore depth manipulation to realize additional hydrodynamic phenomena.

\begin{acknowledgments}
The research of H. Liu was supported by NSFC/RGC Joint Research Scheme, N\_CityU101/21, ANR/RGC Joint Research Scheme, A\_CityU203/19, 
and the Hong Kong RGC General Research Funds (projects 11311122, 11304224 and 11300821).
The research of Z. Miao was supported by the Hong Kong Scholars Program grant XJ2024057.
The research of G. Zheng was supported by the NSF of China (12271151), NSF of Hunan (2020JJ4166) and NSF Innovation Platform Open Fund project of Hunan Province (20K030).
\end{acknowledgments}

\section*{Data Availability Statement}
The data that support the findings of this study are available from
the corresponding authors upon reasonable request.

\appendix
\section{\label{sec:app1}Governing Equations}
The fluid velocity $\tilde{\bm{u}} = (\tilde{\bm{u}}_{\parallel}, \tilde{u}_{\bot}) = (\tilde{u}, \tilde{v}, \tilde{w})$ and pressure $\tilde{p}$ are governed by the continuity and momentum equations for flow in a non-uniform channel
\begin{equation}\label{eq:NS}
  \tilde{\nabla} \cdot (\tilde{h}\tilde{\bm{u}}) = 0, \quad 
  \tilde{\rho} \left( \frac{\partial \tilde{\bm{u}}}{\partial \tilde{t}} + \tilde{\bm{u}} \cdot \tilde{\nabla} \tilde{\bm{u}} \right)
  = -\tilde{\nabla} \tilde{p} + \tilde{\mu} \tilde{\Delta} \tilde{\bm{u}},
\end{equation}
subject to no-slip boundary conditions
\begin{equation}\label{eq:boundary-condition}
    \tilde{\bm{u}}_{\parallel} \big|_{\tilde{z}=0} = \tilde{\bm{u}}_{\parallel} \big|_{\tilde{z}=\tilde{h}} = \bm{0}.
\end{equation}
Here $\tilde{\rho}$, $\tilde{\mu}$, and $\tilde{t}$ denote fluid density, dynamic viscosity, and time, respectively.

We define characteristic scales: pressure $\tilde{p}^{*} = \tilde{\mu}\tilde{u}^{*}\tilde{r}_0/\tilde{h}_0^2$, time $\tilde{t}^{*} = \tilde{r}_0/\tilde{u}^{*}$, and vertical velocity $\tilde{w}^{*} = \epsilon \tilde{u}^{*}$, where $\tilde{u}^{*}$ is the characteristic horizontal velocity, $\tilde{r}_0$ is the characteristic in-plane length, and $\tilde{h}_0$ is the reference channel depth. Introducing normalized variables
\begin{align}
    t &= \frac{\tilde{t}}{\tilde{t}^{*}}, \quad
    x = \frac{\tilde{x}}{\tilde{r}_{0}}, \quad
    y = \frac{\tilde{y}}{\tilde{r}_{0}}, \quad
    z = \frac{\tilde{z}}{\tilde{h}_{0}}, \quad
    \bm{u}_{\parallel} = \frac{\tilde{\bm{u}}_{\parallel}}{\tilde{u}^{*}}, \label{eq:scaling1} \\
    u &= \frac{\tilde{u}}{\tilde{u}^{*}}, \quad
    v = \frac{\tilde{v}}{\tilde{u}^{*}}, \quad
    w = \frac{\tilde{w}}{\tilde{w}^{*}}, \quad
    p = \frac{\tilde{p}}{\tilde{p}^{*}}, \quad
    h = \frac{\tilde{h}}{\tilde{h}_{0}}. \label{eq:scaling2}
\end{align}

For microscale flows, we assume shallow geometry ($\epsilon = \tilde{h}/\tilde{r}_0 \ll 1$) and negligible inertia ($\epsilon \Rey = \epsilon \tilde{\rho}\tilde{u}^{*}\tilde{h}/\tilde{\mu} \ll 1$). Substituting \eqref{eq:scaling1} and \eqref{eq:scaling2} into \eqref{eq:NS} and \eqref{eq:boundary-condition} and retaining leading-order terms yields the lubrication equations
\begin{align}
    \frac{\partial (h u)}{\partial x} + \frac{\partial (h v)}{\partial y} + \frac{\partial (h w)}{\partial z} &= 0, \label{eq:lubrication-continuity} \\
    \nabla_{\parallel} p &= \frac{\partial^2 \bm{u}_{\parallel}}{\partial z^2}, \label{eq:lubrication-momentum} \\
    \frac{\partial p}{\partial z} &= 0, \label{eq:lubrication-boundary-p} \\
    \bm{u}_{\parallel} \big|_{z=0} = \bm{u}_{\parallel} \big|_{z=h} = \bm{0}, \quad
    u_{\bot} \big|_{z=0} &= u_{\bot} \big|_{z=h} = 0, \label{eq:lubrication-boundary-u}
\end{align}
where $\nabla_{\parallel} = (\partial / \partial x, \partial / \partial y)$. From \eqref{eq:lubrication-boundary-p}, the leading-order pressure is $z$-independent, $p = p(x,y)$.

Integrating \eqref{eq:lubrication-momentum} twice over $z$ and applying \eqref{eq:lubrication-boundary-u} yields the depth-averaged velocity
\begin{equation}\label{eq:depth-averaged-u}
    \langle \bm{u}_{\parallel} \rangle = -\frac{h^2}{12} \nabla_{\parallel} p.
\end{equation}
Similarly, depth-averaging \eqref{eq:lubrication-continuity} gives
\begin{equation}\label{eq:depth-averaged-continuity}
    \nabla_{\parallel} \cdot (h \langle \bm{u}_{\parallel} \rangle) = 0.
\end{equation}
Combining \eqref{eq:depth-averaged-u} and \eqref{eq:depth-averaged-continuity} yields the pressure equation
\begin{equation}\label{eq:non-dimension-governing}
    \nabla_{\parallel} \cdot (h^3 \nabla_{\parallel} p) = 0.
\end{equation}

\section{\label{sec:app2}Analytical Solutions}
The general solution to Eq.~\eqref{eq:non-dimension-governing} in polar coordinates $(r,\theta)$ is
\begin{align*}
    p_{\text{in}} &= \left( A r + B r^{-1} \right) \cos\theta,  & r_i &< r < r_e, \\
    p_{\text{out}} &= \left( -12u_{\ext}r + C r^{-1} \right) \cos\theta, & r &> r_e,
\end{align*}
where the subscripts denote pressure fields inside and outside the cloaking region.

The dimensionless boundary conditions are
\begin{align*}
    \left. \frac{\partial p_{\text{in}}}{\partial r} \right|_{r=r_i} &= 0, \\
    p_{\text{in}}|_{r_e} &= p_{\text{out}}|_{r_e}, \\
    \left. h_c^3 \frac{\partial p_{\text{in}}}{\partial r} \right|_{r=r_e} &= \left. \frac{\partial p_{\text{out}}}{\partial r} \right|_{r=r_e},\\
        \lim_{r\to\infty} p_{\text{out}} &= -12u_{\ext}r\cos\theta.
\end{align*}
Solving these yields
\begin{align*}
    A &= -\frac{24r_e^2 u_{\ext}}{(r_e^2 - r_i^2)h_c^3 + (r_e^2 + r_i^2)}, \\
    B &= r_i^2 A, \\
    C &= 12r_e^2 u_{\ext} \frac{(r_e^2 - r_i^2)h_c^3 - (r_e^2 + r_i^2)}{(r_e^2 - r_i^2)h_c^3 + (r_e^2 + r_i^2)}.
\end{align*}
The pressure distribution is therefore
\begin{equation}\label{eq:pressure-circle}
p = \begin{cases}
    \displaystyle 
    -\frac{24r_e^2 u_{\ext}}{(r_e^2 - r_i^2)h_c^3 + (r_e^2 + r_i^2)} \left(r + \frac{r_i^2}{r}\right) \cos\theta, 
    & r_i < r < r_e, \\[2ex]
    \displaystyle 
    12u_{\ext} \left[ \frac{(r_e^2 - r_i^2)h_c^3 - (r_e^2 + r_i^2)}{(r_e^2 - r_i^2)h_c^3 + (r_e^2 + r_i^2)} \frac{r_e^2}{r} - r \right] \cos\theta, 
    & r > r_e.
\end{cases}
\end{equation}
From \eqref{eq:pressure-circle}, the outer flow satisfies cloaking condition \eqref{eq:far-field} when
\begin{equation*}
    h_c = \sqrt[3]{\frac{r_e^2 + r_i^2}{r_e^2 - r_i^2}}.
\end{equation*}

For confocal elliptical cloak, the pressure distribution is
\begin{equation}\label{eq:pressure-ellipse}
p = \begin{cases}
    \displaystyle 
    -\frac{12\re^{2\xi_e} u_{\ext}}{(\re^{2\xi_e} - \re^{2\xi_i}) h_c^3 + (\re^{2\xi_e} + \re^{2\xi_i})} \left(\re^{\xi} + \re^{2\xi_i - \xi}\right) \cos\eta, \\
     \quad \xi_i < \xi < \xi_e, \\[2ex]
    \displaystyle 
    12u_{\ext} \bigg[ \frac{\cosh\xi_e (\re^{2\xi_e} - \re^{2\xi_i}) h_c^3 - \sinh\xi_e (\re^{2\xi_e} + \re^{2\xi_i}) }{(\re^{2\xi_e} - \re^{2\xi_i}) h_c^3 + (\re^{2\xi_e} + \re^{2\xi_i})} \frac{\re^{\xi_e}}{\re^\xi}\\
     - \cosh\xi \bigg] \cos\eta, \quad
     \xi > \xi_e.
\end{cases}
\end{equation}
From \eqref{eq:pressure-ellipse}, cloaking condition \eqref{eq:far-field} is satisfied when
\begin{equation*}
    h_c = \sqrt[3]{\tanh\xi_e \coth(\xi_e - \xi_i)}.
\end{equation*}

\nocite{*}
\bibliography{reference}

\end{document}